\documentclass[reprint,superscriptaddress,amsmath,amssymb,aps,floatfix]{revtex4-1}
\usepackage[utf8]{inputenc}
\usepackage{graphicx}
\usepackage{dcolumn}
\usepackage{bm}
\usepackage[colorlinks,citecolor=blue,linkcolor=blue,urlcolor=blue]{hyperref}
\usepackage{booktabs}
\usepackage{float}

\begin{document}

\title{Ab Initio Studies on Interactions in K$_3$C$_{60}$ under High Pressure}

\author{Jianyu Li}
\affiliation{School of Science, Harbin Institute of Technology, Shenzhen, 518055, China}

\author{Zhangkai Cao}
\affiliation{School of Science, Harbin Institute of Technology, Shenzhen, 518055, China}

\author{Jiahao Su}
\affiliation{School of Science, Harbin Institute of Technology, Shenzhen, 518055, China}

\author{Ruipeng Wang}
\affiliation{School of Materials Science and Physics, China University of Mining and Technology, Xuzhou, 221116, China}

\author{Haipeng Li}
\affiliation{School of Materials Science and Physics, China University of Mining and Technology, Xuzhou, 221116, China}

\author{Yusuke Nomura}
\affiliation{Department of Applied Physics and Physico-Informatics, Keio University, Yokohama 223-8522, Japan }

\author{Xiaosen Yang}
\email{yangxs@ujs.edu.cn}
\affiliation{Department of Physics, Jiangsu University, Zhenjiang, 212013, China.}

\author{Ho-Kin Tang}
\email{denghaojian@hit.edu.cn}
\affiliation{School of Science, Harbin Institute of Technology, Shenzhen, 518055, China}

\date{\today}

\begin{abstract}
Fullerene solids doped with alkali metals (A$_3$C$_{60}$, A = K, Rb, Cs) exhibit a superconducting transition temperature ($T_c$) as high as 40 K, and their unconventional superconducting properties have been a subject of debate.
With application of high pressure on K$_3$C$_{60}$ and Rb$_3$C$_{60}$, the experiments demonstrate the decrease of $T_c$.
In this paper, we focus on K$_3$C$_{60}$ and derive the structure of K$_3$C$_{60}$ under different pressures based on first-principles calculations, exploring the trends of Coulomb interactions at various pressures.
By utilizing the Maximally Localized Wannier function approach, Constrained Density Functional Perturbation Theory (cDFPT), and Constrained Random Phase Approximation (cRPA), we construct a microscopic low-energy model near the Fermi level. 
Our results strongly indicate that, in the K$_3$C$_{60}$ system, as pressure increases, the effect of phonons is the key to intraorbital  electron pairing. 
There is a dominance of the phonon-driven superconducting mechanism at high pressure.
 
\end{abstract}
\pacs{Valid PACS appear here}

\maketitle
\section{Introduction} 

Since their discovery three decades ago, fullerenes have commanded substantial scientific attention~\cite{Stankevich1984-zb,Smalley1997-ay,Wang2022-jd,Wang2023-ng}. Notably, the intriguing hallmark of these molecules resides in their distinctive geometric structure: C$_{60}$ molecules exhibit the symmetry of an icosahedral group, a symmetry class of paramount significance within the realm of symmetric operations. The recent realization of monolayer fullerene materials~\cite{Hou2022-yg,Yu2023-hv,Tromer2022-gl,Yu2022-hx,Peng2022-hq,Ying2023-lt,Dong2023-rn,Yuan2023-xc} further underscores the importance of investigating fullerene materials' physical properties, as these insights are pivotal in unlocking their full technological potential.

After successfully synthesizing solid C$_{60}$ (fullerene crystals), a series of extensive solid-state experiments has been conducted, revealing numerous significant physical properties~\cite{Xiang1992-ts,Klein1992-tg,Hou1995-um,Hebard1993-ek}. For instance, fcc K$_3$C$_{60}$ demonstrates metallic behavior, with the resistivity increasing as the temperature rises~\cite{Xiang1992-ts}. 
A distinct Drude peak is observed in the optical conductivity, and photoemission studies have identified a finite density of states near the Fermi level~\cite{Chen1991-lc,Benning1993-wj,Hesper2000-js,Yang2003-wr}.
Pressure tuning of light-induced superconductivity in K$_3$C$_{60}$ has also made significant progress~\cite{Cantaluppi2018-sr,Rowe2023-ti}.
Measurements of electron spin resonance~\cite{Tanigaki1995-vk,Robert1998-eg} and nuclear magnetic resonance~\cite{Tycko1992-fk,Stenger1995-fj,Pennington1996-io} indicate a Pauli-like susceptibility behavior in the normal state, which is nearly temperature-independent.

A groundbreaking achievement was made by Hebard et al. as they successfully identified superconductivity in potassium-doped C$_{60}$, notably characterizing a superconducting transition at a critical temperature of $T_{c}$ = 18 K~\cite{Hebard1991-xg}. Subsequent investigations into the underlying superconducting mechanisms have revealed resemblances to the Bardeen-Cooper-Schrieffer (BCS) theory, supported by consistent experiments. These include the confirmation of a full-gap $s$-wave pairing~\cite{Zhang1991-ee,Zhang1991-lo,Tycko1992-fk,Kiefl1993-uq}, reduction in spin susceptibility within the superconducting phase~\cite{Stenger1995-fj}, observation of the Hebel-Slichter peak~\cite{hebel1959nuclear} in nuclear magnetic resonance (NMR)~\cite{Sasaki1994-nr,Stenger1995-fj}, and the correlation between lattice constant and $T_{c}$~\cite{Fleming1991-ox}. 
 
Moreover, there have been thorough examinations of pressure-induced alterations in $T_{c}$~\cite{Sparn1991-kz,Schirber1991-ma,Wang2022-wj} and lattice parameter~\cite{Zhou1992-vt,Wang2022-wj} within the K$_3$C$_{60}$ system. 
In non-Cs-doped C$_{60}$ superconductors, the observed fact of $T_c$ monotonically decreasing under lattice compression has been attributed to the band widening caused by lattice contraction, resulting in a reduction of the density of states at the Fermi level~\cite{Rosseinsky1991-dx,Fleming1991-ox,Sparn1991-kz,Zhou1992-vt,Tanigaki1992-ct,Yildirim1995-vu}. 
In the context of the general reduction in $T_{c}$ under pressure, these experimental findings provide the conditions for further exploration of the superconducting mechanism in K$_3$C$_{60}$.

Given the intricate interplay between electronic correlations and complex electron-phonon coupling (EPC), the mechanisms underpinning superconductivity in fullerenes remain contentious. While initial perceptions categorized fullerene superconductors as conventional BCS systems, the discourse has expanded to encompass an array of unconventional theories. These include propositions of polaron-driven superconductivity~\cite{Takahashi1992-cb,Tiwari2009-gn}, local pairing facilitated by Jahn-Teller phonons and Coulomb repulsion~\cite{Han2003-ds,Nomura2016-lk}, the intriguing concept of negative Hund's coupling stabilized by EPC~\cite{Capone2002-sy,Nomura2015-kz}, and even scenarios of pure electronic pairing~\cite{Jiang2016-kg}, among other possibilities.

In this context, our present study initiates an $ab$ $initio$ exploration of interactions within the K$_3$C$_{60}$ system under the influence of elevated pressures. Utilizing experimentally acquired data on lattice constants under various pressures~\cite{Wang2022-wj}, we aim to employ a combination of Density Functional Theory (DFT) and the model-calculation approach~\cite{Kotliar2006-av,Held2007-yp,Imada2010-qt} to compute essential interaction parameters.
We will determine the effective Coulomb interaction within the low-energy subspace using the constrained random phase approximation (cRPA)~\cite{Aryasetiawan2004-iz}. Additionally, we will calculate partially renormalized phonon frequencies and electron–phonon couplings through Constrained Density Functional Perturbation Theory (cDFPT)~\cite{Nomura2014-mz,PhysRevB.92.245108}.
Our primary objective is to unveil the alterations in physical properties, ranging from electronic characteristics to phonon behavior, caused by pressure fluctuations.
By closely examining the pressure-induced modifications in various interaction parameters, we strive to discern the most plausible superconducting mechanism amid the changing pressure conditions.

The rest of this paper is outlined as follows. In Sec.\ \ref{Method and Model}, we choose the target band and show the low-energy models from $ab$ $initio$ calculations. 
In Sec.\ \ref{RESULT}, we calculate the global band structure using DFT and construct MLWO’s within the low-energy band subspace. Detailed computations are presented in this section. 
We obtained parameter results for the Coulomb interaction and electron-phonon interaction and analyzed their trends with varying pressure.
In Sec.\ \ref{discussion}, we discuss changes in electronic and phononic properties resulting from pressure fluctuations and compare various explanations for the unconventional superconducting mechanism in K$_3$C$_{60}$.
Finally, we present the conclusion in Sec.\ \ref{Conlusion}.

\section{Method and Model}
\label{Method and Model}

\begin{figure}[b!]
\includegraphics[width=1\columnwidth]{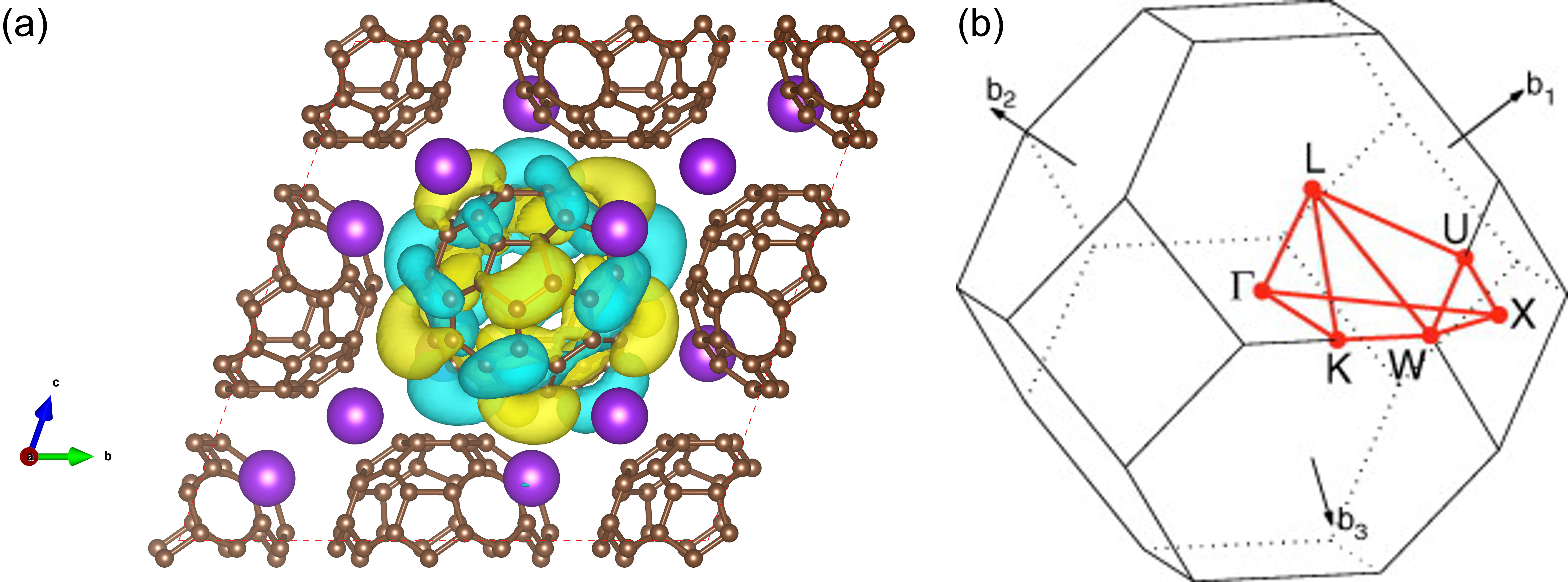}
    \caption{(Color online) (a) illustrates the atomic structure of K$_3$C$_{60}$, which is set at a size of $2 \times 2 \times 2 $, and the real-space Wannier function displayed using VESTA software~\cite{VESTA}. K atoms are depicted as purple spheres, while C atoms are represented by brown spheres. Positive isosurfaces are highlighted in yellow, while negative isosurfaces are shown in blue. (b) Brillouin zone of the face-centered cubic (fcc) lattice, path: $\Gamma -X-U-K-\Gamma -L-K-W-X$ as described in Ref.~\cite{Setyawan2010-uy}.}
\label{supercell}
\end{figure}

Close to the Fermi energy, $t_{1u}$ band is the focus of our investigation which we call target band. We use a lattice Hamiltonian encompassing the electron and phonon degrees of freedom specifically associated with the $t_{1u}$ orbitals that  has been formulated in Refs.~\cite{PhysRevB.85.155452,Nomura2015-kz}. Nearly all the excitation processes occur in $t$-subspace,
the subspace which the target bands span (For later use, we define $r$-subspace as the rest of
the Hilbert space). 
The Hamiltonian assumes the following general form:

\begin{align}
\mathcal{H} &= \sum_{ij\mathbf{k}\sigma}{\mathcal{H} _{ij}^{(0)}}(\mathbf{k})c_{i\mathbf{k}}^{\sigma \dagger}c_{j\mathbf{k}}^{\sigma} \notag \\
&\quad + \sum_{\mathbf{qkk}^{\prime}iji^{\prime}j^{\prime}}{\sum_{\sigma \sigma ^{\prime}}{U_{ij,i^{\prime}j^{\prime}}}}(\mathbf{q})c_{i\mathbf{k}+\mathbf{q}}^{\sigma \dagger}c_{j^{\prime}\mathbf{k}^{\prime}}^{\sigma ^{\prime}\dagger}c_{i^{\prime}\mathbf{k}^{\prime}+\mathbf{q}}^{\sigma ^{\prime}}c_{j\mathbf{k}}^{\sigma} \notag \\
&\quad + \sum_{ij\mathbf{k}\sigma \mathbf{q}\nu }{\sum_{ij}{g_{ij}^{\nu}(}}\mathbf{k},\mathbf{q})c_{i\mathbf{k}+\mathbf{q}}^{\sigma \dagger}c_{j\mathbf{k}}^{\sigma}(b_{\mathbf{q}\nu }+b_{-\mathbf{q}\nu }^{\dagger}) \notag \\
&\quad + \sum_{\mathbf{q}\nu }{\omega _{\mathbf{q}\nu }}b_{\mathbf{q}\nu }^{\dagger}b_{\mathbf{q}\nu },
\label{Hamiltonian}
\end{align}

This Hamiltonian operates on the fcc lattice, where each lattice site corresponds to a C$_{60}$ molecule. The terms $\mathcal{H}^{(0)}$, $U$, $g$, and $\omega$ respectively denote the one-body electron Hamiltonian, Coulomb interaction, electron-phonon coupling, and phonon frequency. All these quantities are indexed by Wannier orbitals $i$, $j$, $i^{\prime}$, $j^{\prime}$, spin $\sigma$, $\sigma^{\prime}$, momentum $\mathbf{k}$, $\mathbf{q}$, and the phonons are further distinguished by a branch index $\nu$.

To ensure the short-ranged nature of hopping, we have constructed maximally localized Wannier orbitals~\cite{wannier} as the basis (Fig.~\ref{supercell}(a)). These parameters are determined using state-of-the-art $ab$ $initio$ techniques, with particular emphasis on the utilization of cRPA~\cite{Aryasetiawan2004-iz} for the calculation of electron-related terms $U_{ij,i^{\prime}j^{\prime}}$
and the cDFPT~\cite{Nomura2014-mz} for the calculation of phonon-related terms $g_{i j}^\nu$ and $\omega_{\mathbf{q} \nu}$, wherein the influence of high-energy bands is incorporated into the parameter values.

The parameters $\mathcal{H} _{ij}^{(0)}(\mathbf{k})
$ represent an on-site energy $\left(\bf{k}=\mathbf{0}\right)$ and hopping integrals $\left(\bf{k} \neq \mathbf{0}\right)$, which are described with the translational symmetry as 
\begin{align}
    \mathcal{H} _{ij}^{(0)}(\mathbf{k})=\left< \phi _{i\mathbf{k}} \right|\mathcal{H} _{\mathrm{KS}}\left| \phi _{j\mathbf{0}} \right> ,
\end{align}
where $|\phi _{i\mathbf{k}}\rangle =c_{i\mathbf{k}}^{\dagger}|0\rangle $ and $\mathcal{H} _{\mathrm{KS}}$ is the Kohn-Sham Hamilton.

To evaluate effective interaction parameters $U_{ij,i^{\prime}j^{\prime}}(\mathbf{q})$, 
we convert $U_{ij,i^{\prime}j^{\prime}}(\mathbf{q})$ into the onsite Coulomb repulsion $U_{i j}$ and the exchange interaction $J_{i j}$
and calculate the screened Coulomb interaction $W$ at the low-frequency limit. We first calculate the noninteracting-polarization function $\chi$, excluding polarization processes within the target bands. Note that screening by the target electrons is considered when we solve the effective models so that we have to avoid double counting of it when we derive the effective models. With the resulting $\chi$, the $W$ interaction is calculated as $W=(1-v \chi)^{-1} v$, where $v$ is the bare Coulomb interaction potential $v\left(\mathbf{r}, \mathbf{r}^{\prime}\right)=\frac{1}{\left|\mathbf{r}-\mathbf{r}^{\prime}\right|}$.
 We consider a decomposition of the total irreducible polarization $\chi^0$ into the one involving only the $t$-subspace electrons $\chi_t^0$ and the rest $\chi_r^0$~\cite{Aryasetiawan2004-iz} :
$$
\chi^0=\chi_t^0+\chi_r^0 .
$$
Note that $\chi_r^0$ contains not only the processes involving only the $r$-subspace degrees of freedom but also the ones involving both the $t$-subspace electrons and the $r$-subspace electrons. We define the partially screened Coulomb interaction $W^{(p)}$ as
\begin{align}
    W^{(p)}=\left(1-v \chi_r^0\right)^{-1} v=\epsilon_r^{-1} v
\end{align}
The onsite Coulomb repulsion $U_{i j}$ and the exchange interaction $J_{i j}$ are evaluated as
\begin{align}
U_{i j} & =\iint d \mathbf{r} d \mathbf{r}^{\prime}\left|\phi_{i \mathbf{0}}(\mathbf{r})\right|^2 W^{(p)}\left(\mathbf{r}, \mathbf{r}^{\prime}\right)\left|\phi_{j \mathbf{0}}\left(\mathbf{r}^{\prime}\right)\right|^2 \notag\\
& =\frac{4 \pi e^2}{N \Omega} \sum_{\mathbf{q}} \sum_{\mathbf{G} \mathbf{G}^{\prime}} \rho_{i i}(\mathbf{q}+\mathbf{G}) W_{\mathbf{G}, \mathbf{G}^{\prime}}^{(p)}(\mathbf{q}) \rho_{j j}^*\left(\mathbf{q}+\mathbf{G}^{\prime}\right)
\label{U}
\end{align}
and
\begin{align}
J_{i j} & =\iint d \mathbf{r} d \mathbf{r}^{\prime} \phi_{i \mathbf{0}}^*(\mathbf{r}) \phi_{j \mathbf{0}}(\mathbf{r}) W^{(p)}\left(\mathbf{r}, \mathbf{r}^{\prime}\right) \phi_{j \mathbf{0}}^*\left(\mathbf{r}^{\prime}\right) \phi_{i \mathbf{0}}\left(\mathbf{r}^{\prime}\right) \notag\\
& =\frac{4 \pi e^2}{N \Omega} \sum_{\mathbf{q}} \sum_{\mathbf{G G}^{\prime}} \rho_{i j}(\mathbf{q}+\mathbf{G}) W_{\mathbf{G}, \mathbf{G}^{\prime}}^{(p)}(\mathbf{q}) \rho_{i j}^*\left(\mathbf{q}+\mathbf{G}^{\prime}\right),
\label{J}
\end{align}
respectively, where $\Omega$ is the volume of the unit cell, and $\rho_{i j}(\mathbf{q}+\mathbf{G})$ is given, with the Wannier-gauge Bloch functions $\psi_{i \mathrm{k}}^{(w)}$, by
\begin{align}
    \rho_{i j}(\mathbf{q}+\mathbf{G})=\frac{1}{N} \sum_{\mathbf{k}}\left\langle\psi_{i \mathbf{k}+\mathbf{q}}^{(w)}\left|e^{i(\mathbf{q}+\mathbf{G}) \cdot \mathbf{r}}\right| \psi_{j \mathbf{k}}^{(w)}\right\rangle .
\end{align}


To facilitate a comparative analysis with the cRPA results, we compute interaction parameters using the unscreened
case, representing the bare Coulomb interaction. In order to differentiate it from the cRPA results, we refer to this as the  ``bare'' interaction.

In solids, the individual ion can be identified as the $\kappa$-th ion in the $p$-th unit cell and the direction of the displacement $\alpha$ = $x$, $y$, $z$.
To evaluate effective electron-phonon-coupling interaction parameters, we use
\begin{align}
    g_{ij}^{\nu}(\mathbf{k},\mathbf{q})=\sum_{\kappa \alpha}{\sqrt{\frac{\hbar}{2M_{\kappa}\omega _{\mathbf{q}\nu}}}}e_{\kappa}^{\left( p \right) \alpha}(\mathbf{q}\nu)\left. \langle \phi _{i\mathbf{k}+\mathbf{q}}\left| \frac{\partial V^{\left( p \right)}(\mathbf{r})}{\partial u_{\kappa}^{\alpha}(\mathbf{q})} \right|\phi _{j\mathbf{k}} \right. \rangle ,
\end{align}
where we employ the Wannier-gauge for the electrons, and the superscript $(p)$ indicates the partially renormalized quantities. 

Here, we consider the real-phonon contributions the intraorbital density-density-type interaction $U_{\rm ph}$, interorbital density-density-type interaction $U_{\rm ph}^{\prime}$ and exchange-type
interaction $J_{\rm ph}$ :
\begin{align}
	U_{\rm ph}&=-\frac{1}{N_{\mathbf{q}}}\sum_{\mathbf{q}\nu}{\frac{2\left| \tilde{g}_{ii}^{(p)}(\mathbf{q},\nu) \right|^2}{\omega _{\mathbf{q}\nu}^{(p)}},}\\
	U_{\rm ph}^{\prime}&=-\frac{1}{N_{\mathbf{q}}}\sum_{\mathbf{q}\nu}{\frac{2\tilde{g}_{ii}^{(p)}(\mathbf{q},\nu)\tilde{g}_{jj}^{(p)*}(\mathbf{q},\nu)}{\omega _{\mathbf{q}\nu}^{(p)}},}\\
	J_{\rm ph}&=-\frac{1}{N_{\mathbf{q}}}\sum_{\mathbf{q}\nu}{\frac{2\tilde{g}_{ij}^{(p)}(\mathbf{q},\nu)\tilde{g}_{ij}^{(p)*}(\mathbf{q},\nu)}{\omega _{\mathbf{q}\nu}^{(p)}}}\notag\\
	&=-\frac{1}{N_{\mathbf{q}}}\sum_{\mathbf{q}\nu}{\frac{2\tilde{g}_{ij}^{(p)}(\mathbf{q},\nu)\tilde{g}_{ji}^{(p)*}(\mathbf{q},\nu)}{\omega _{\mathbf{q}\nu}^{(p)}},}
\end{align}
where the orbital dependences of $U_{\rm ph}$, $U_{\rm ph}^{\prime}$ and $J_{\rm ph}$ do not exist by symmetry, and $\tilde{g}_{i j}^{(p)}(\mathbf{q}, \nu)$ is calculated as
\begin{align}
  \tilde{g}_{i j}^{(p)}(\mathbf{q}, \nu)=\frac{1}{N_{\mathbf{k}}} \sum_{\mathbf{k}} g_{i j}^{(p) \nu}(\mathbf{k}, \mathbf{q})  
\end{align}

\section{RESULT}
\label{RESULT}
\subsection{Calculation details}
We construct maximally localized Wannier orbitals (MLWOs) (Fig.~\ref{supercell}) from the DFT $t_{1 u}$ band, and calculate the model parameters in the lattice Hamiltonian [Eq.~(\ref{Hamiltonian})] as described in Methods.
We employed the Local Density Approximation (LDA) exchange-correlation functional with the parameterization by Perdew-Burke-Ernzerhof and employed the Troullier-Martins conservative pseudopotentials within the Kleinman-Bylander representation. The configurations used to generate the pseudopotentials for C and K,  are $(2s)^{2}(2p)^{2}$ and $(3 p)^{6}(4 s)^{0}(3 d)^{0}$, respectively. A partial core correction was applied to the pseudopotential for the alkali metal K. The unit cell volume data for K$_3$C$_{60}$ at different pressures were obtained from experimental sources~\cite{Wang2022-wj}. Calculations were performed using the open-source software Quantum Espresso~\cite{qe1,qe2,qe3} and RESPACK~\cite{respack1,respack2,respack3,respack4,respack5,respack6}. We perform the structure optimization with fixing the lattice constants to the values employed~\cite{Wang2022-wj} and ignoring the orientational disorder of C$_{60}$ molecules. 

In DFT calculations, crystal lattice is defined as a type 2 Bravais lattice, with the lattice parameter $a$ determined from Fig.~\ref{lattice constant}(a). The calculation includes 350 electronic bands. Electron occupation numbers are determined using the "smoothing" method with a Gaussian smoothing of 0.025 Ry.
In our Wannier orbital calculations, we have specified the following parameters: a crystal cell with a multiplicity of 2, a computation for 3 Wannier orbitals, and an energy window ranging from $E_{\mathrm{F}}-$0.5 to $E_{\mathrm{F}}+$0.5 eV ($E_{\mathrm{F}}$ is Fermi energy). We have employed a set of 6 initial Gaussian functions as a foundational basis. Moreover, we have defined Gaussian function coefficient matrices for distinct orbital types, such as $p_x$, $p_y$, and $p_z$. For example, in the case of the $p_x$ orbital, the coefficient matrix comprises four values: 1.0, $-$0.243, 0.243, and 0.243, which represent the weights assigned to each Gaussian function utilized in constructing the $p_x$ orbital. Analogously, $p_y$ and $p_z$ orbitals possess analogous coefficient matrices for specifying their Gaussian function weights. Additionally, we have established a path for symmetric $\mathbf{k}$-point interpolation, encompassing specific high-symmetry points:  
$\Gamma$ (0, 0, 0), $X$ (0.5, 0, 0.5), $U$ (0.625, 0.25, 0.625), $K$ (0.375, 0.375, 0.75), $L$ (0.5, 0.5, 0.5), $W$ (0.5, 0.25, 0.75) (Fig.~\ref{supercell}(b)).
Then, we calculate the band structure for the optimized structure.

In order to maximize the utilization of computational resources, different cutoff energies for wave functions and charge densities, as well as $\mathbf{k}$-point sampling grids, were set because the calculation requirements vary when computing Coulomb interactions using the cRPA method and electron-phonon interactions using the cDFPT method.
In the calculation of effective parameters for Coulomb interaction, the wave function and charge density were truncated with energy cutoffs of 36 Ry and 144 Ry, respectively. A $\mathbf{k}$-point sampling grid of $5 \times 5 \times 5$ $\mathbf{k}$-points was utilized.
But in the calculation of effective parameters for electron-phonon interactions, the cutoff energy for the wave functions is set to be 50 Ry, and we employ $4 \times 4 \times 4$ $\mathbf{k}$-points. 
The cDFPT calculations are performed with $2 \times 2 \times 2 $ $\mathbf{q}$-mesh.

\subsection{Band structure}

\begin{figure*}[htb!]
\begin{center}
    \centering    \includegraphics[width=2\columnwidth]{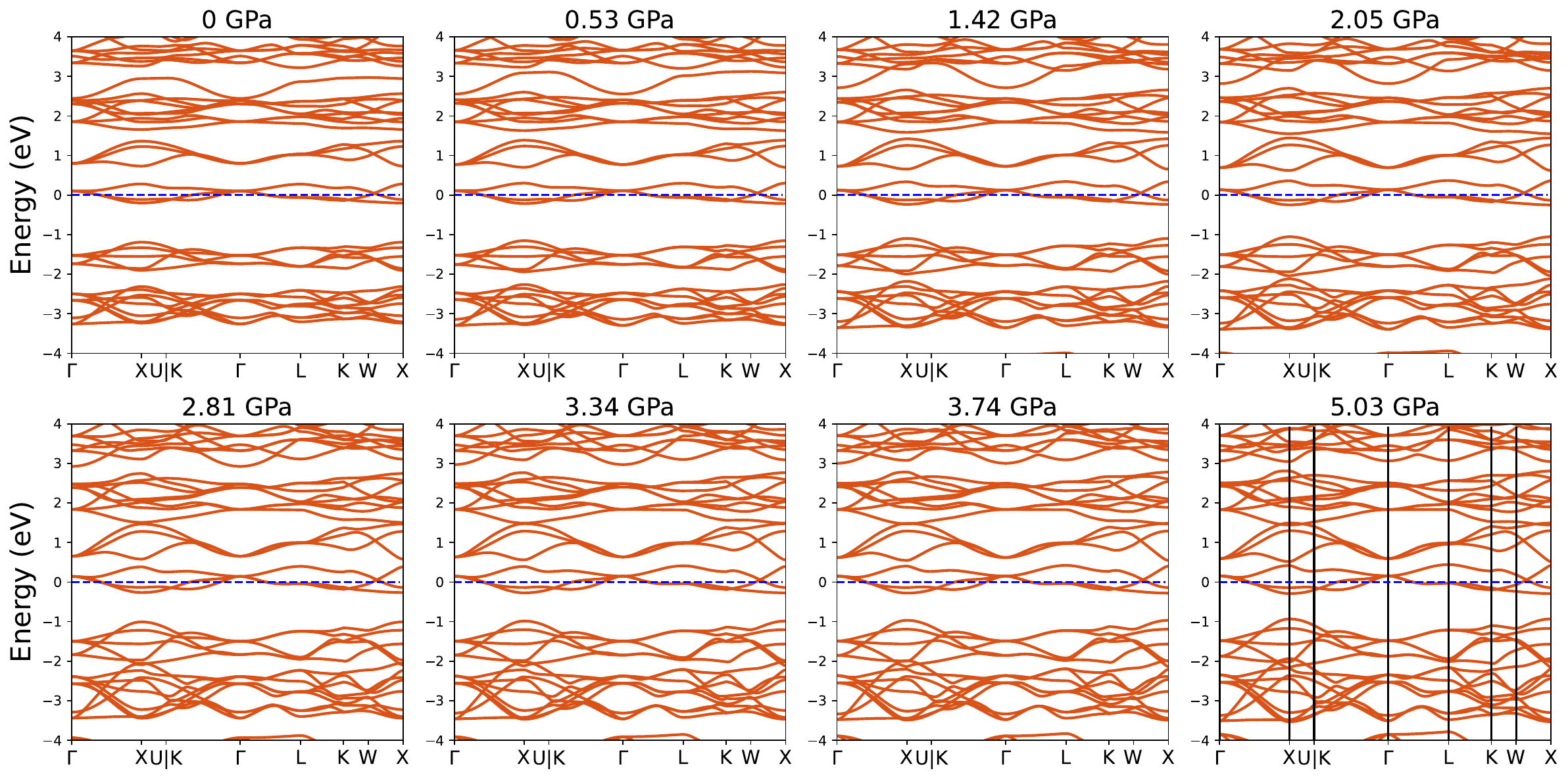}
    \caption{(Color online) Calculated $ab$ $initio$ electronic band structure of fcc-K$_3$C$_{60}$ at different pressures. The pressure parameters are set as 0, 0.53, 1.42, 2.05, 2.81, 3.34, 3.74, 5.03 GPa. The horizontal axis is labeled with the special points in the Brillouin zone: 
    $\Gamma$ (0, 0, 0), $X$ (0.5, 0, 0.5), $U$ (0.625, 0.25, 0.625), $K$ (0.375, 0.375, 0.75), $L$ (0.5, 0.5, 0.5), $W$ (0.5, 0.25, 0.75). Fermi energy is depicted as blue dashed lines.}
\label{band}    
\end{center}
\end{figure*}

\begin{figure}[htb!]
\includegraphics[width=1\columnwidth]{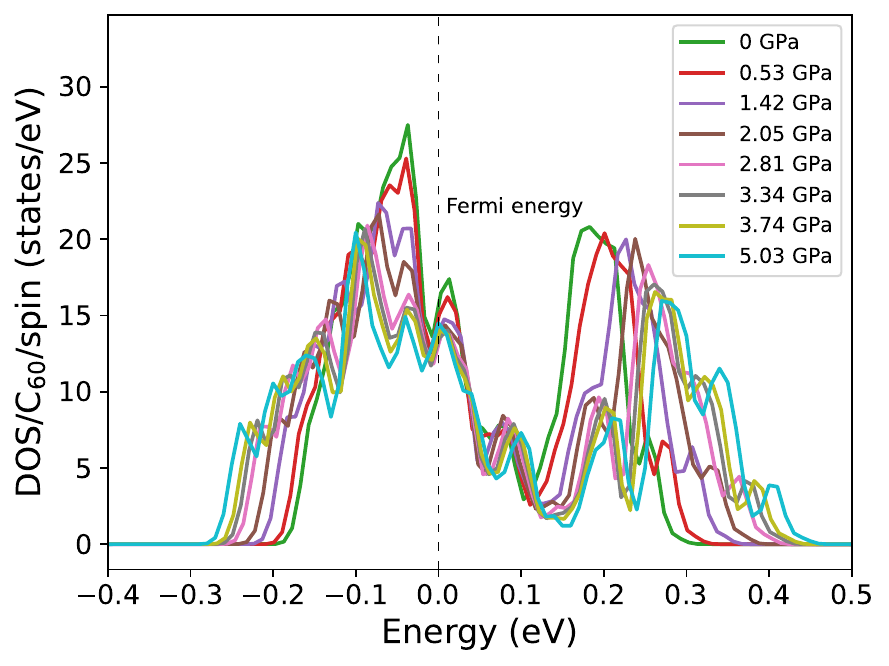}
    \caption{(Color online) Our calculated density of states (DOS) for the $t_{1u}$ band of fcc-K$_3$C$_{60}$ at different pressures. The pressure parameters are set as 0, 0.53, 1.42, 2.05, 2.81, 3.34, 3.74, 5.03 GPa.}
\label{dos}
\end{figure}

\begin{figure}[b!] 
\includegraphics[width=\columnwidth]{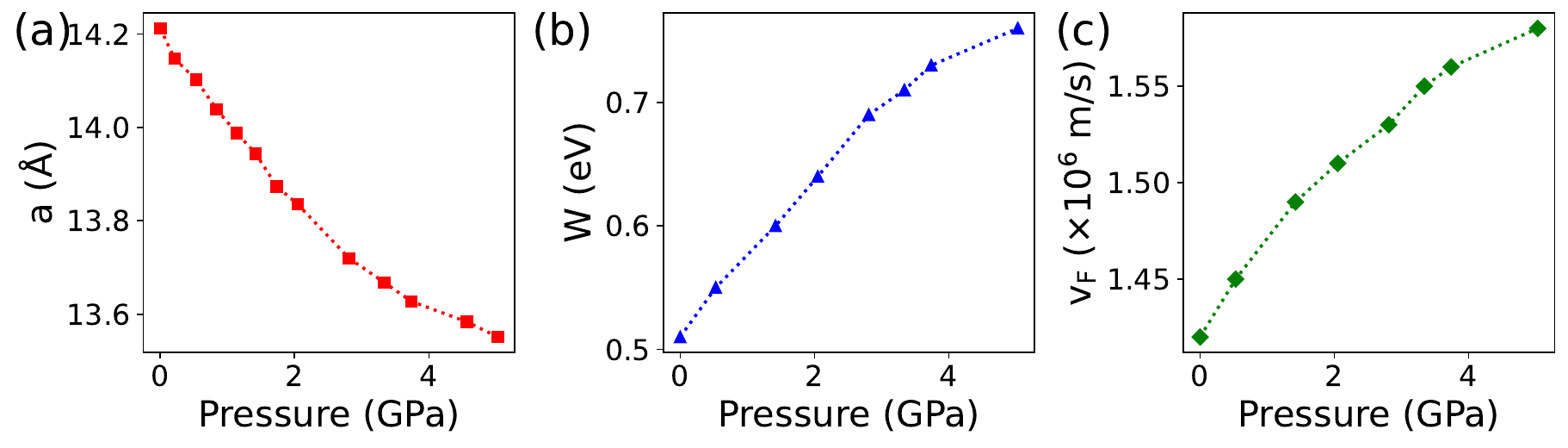}
    \caption{(Color online) The pressure dependence of (a) lattice constant $a$, (b) bandwidth $W$ and (c) Fermi velocity $v_{\mathrm{F}}$. We fix the lattice constants to the experimental values~\cite{Wang2022-wj} and calculate the bandwidth of the $t_{1u}$ band by Quantum Espresso~\cite{qe1,qe2,qe3}. We combine Eq.~(\ref{vf}) to obtain the Fermi velocity from the energy band results.}
\label{lattice constant}
\end{figure}


The C$_{60}$ solid possesses a three-dimensional crystal structure~\cite{c601,c602} owing to the relatively isotropic shape of the C$_{60}$ molecule. The centers of these molecules form an fcc lattice. 
Considering that undoped C$_{60}$ constitutes a molecular solid, an insightful initial approach to comprehend its electronic structure is to examine the molecular limit. In this limit, numerous molecular orbitals degenerate due to the high symmetry inherent to the molecule itself: specifically, the C$_{60}$ molecule boasts the fivefold-degenerate HOMO, threefold-degenerate LUMO, and threefold degenerate LUMO+1 orbitals, denoted as $h_u$, $t_{1 u}$ and $t_{1 g}$ orbitals, respectively, based on their symmetry properties. In the context of a solid, these orbitals acquire a dispersion as a result of electron transfers taking place between the molecules. However, owing to the relatively small magnitude of the transfer integral, the bandwidth of each orbital remains narrow (typically around $\sim 0.5 $ $\rm{eV}$). Consequently, there is minimal overlap observed between the bands associated with different molecular orbitals. Meanwhile, the C$_{60}$ solid manifests as a band insulator in which the LUMO $t_{1 u}$ band remains unoccupied, while the HOMO $h_u$ band is fully populated.

In the fcc K$_3$C$_{60}$, each alkali atom donates about one electron into the $t_{1u}$ band, so the $t_{1u}$ band becomes half-filled. As a result, K$_3$C$_{60}$ is no longer an insulator; instead, it exhibits partial metallic characteristics.
Fig.~\ref{band} illustrates the computed band structures of fcc $\mathrm{K}_3 \mathrm{C}_{60}$ at various pressure levels within the framework of LDA. These compounds exhibit shared characteristics in their band structures, particularly notable are the presence of narrow bands in proximity to the Fermi level, which are well-isolated from other bands. 
Within the K$_3$C$_{60}$, there exists a set of threefold-degenerate states, known as the $t_{1u}$ band, which occupies a prominent position in the vicinity of the Fermi level.
This distinctive feature renders them particularly suitable for selection as target bands in the construction of an effective model.

We show in Fig.~\ref{dos} the calculated density of states (DOS)
of the $t_{1u}$ band for fcc-K$_3$C$_{60}$. DOS configuration profile shows only relatively small change, in which the profile generally obeys the DOS behavior of the Fermi gas.
This suggests we can use the Fermi gas model to quantify Fermi parameters. For the Drude model, we define the density of states $N(E)$, Fermi energy $E_{\mathrm{F}}$ and Fermi velocity $v_{\mathrm{F}}$, respectively: 
\begin{align}
N(E)&=\frac{2V}{(2\pi )^2}(\frac{2m^*}{\hbar ^2})^{3/2}E^{1/2},
\\
E_{\mathrm{F}}&=\hbar ^2\frac{k_{\mathrm{F}}^{2}}{2m^*},
\\
v_{\mathrm{F}}&=\hbar \frac{k_{\mathrm{F}}}{m^*}.
\label{vf}
\end{align}

We find that $N(E)\propto v_{\mathrm{F}}$ and $E_F\propto v^2_{\mathrm{F}}$, demonstrating a consistent alignment between theoretical and computational trends. Increasing pressure causes a reduction in the lattice constant of K$_3$C$_{60}$ (Fig.\ \ref{lattice constant}(a)), leading to an upward shift of the Fermi level and an increase in the Fermi velocity. Additionally, alterations in lattice constant impact the electronic band structure, resulting in a wider bandwidth and influencing the density of states. Both bandwidth $W$ and Fermi velocity $v_{\mathrm{F}}$ exhibit monotonic increase as the lattice constant decreases , $W$ rises from 0.51 eV to 0.76 eV (Fig.~\ref{lattice constant}(b)), representing an approximate 50\% increase, and $v_{\mathrm{F}}$ rises from 1.42 $\times 10^6$m/s to 1.58 $\times 10^6$m/s (Fig.~\ref{lattice constant}(c)), representing an approximate 11\% increase.

\subsection{Coulomb interaction parameters of K$_3$C$_{60}$}

\begin{figure}[b!]   
\includegraphics[width=1\columnwidth]{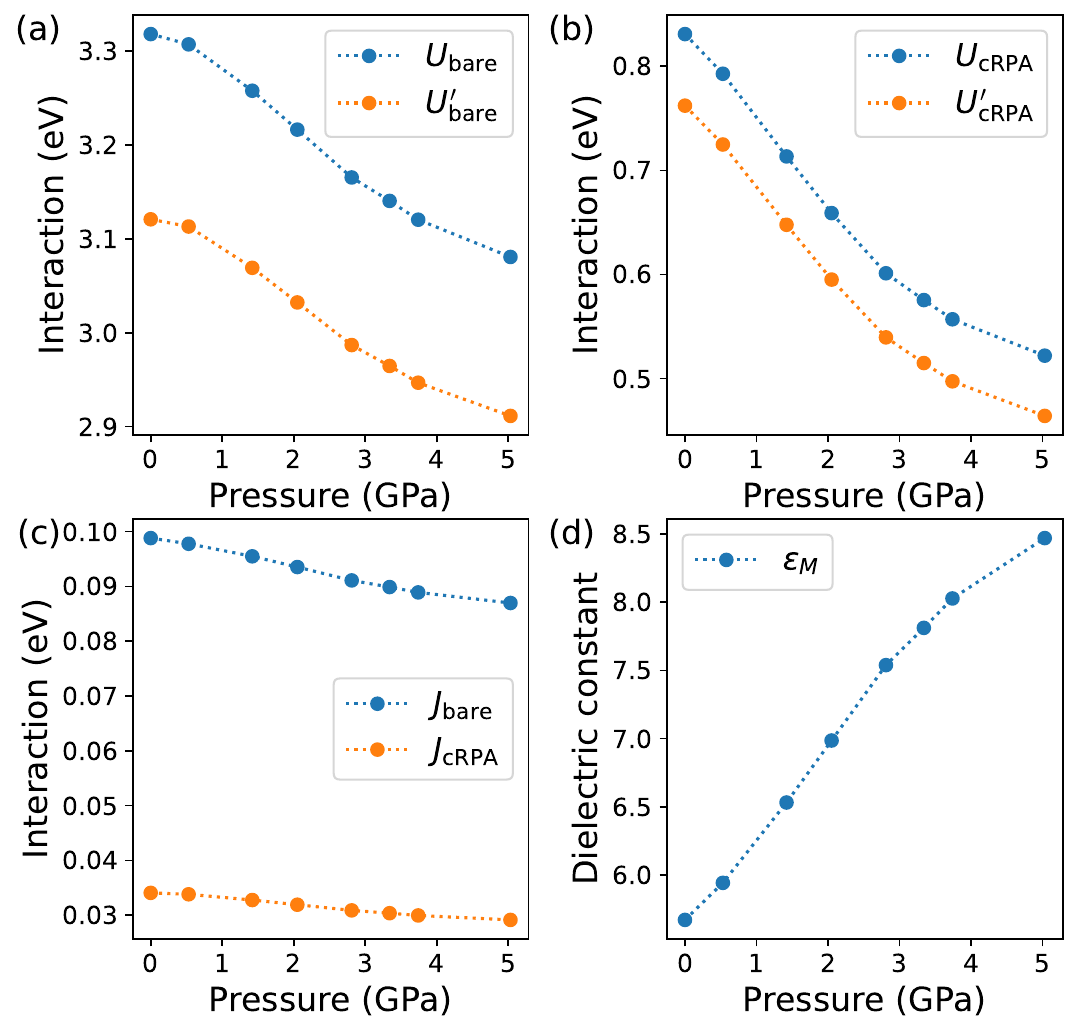}
    \caption{(Color online) (a)-(c) $U$, $U^{\prime}$ and $J$ with different screening levels [unscreened (bare) and constrained RPA (cRPA)].
    For bare and cRPA $U$, $U^{\prime}$ and $J$ values, the units are given in eV. (d) $\epsilon _{M}$ is the cRPA-macroscopic-dielectric constant in Eq.~(\ref{epsilon}).}
\label{cRPA}
\end{figure}

\begin{figure}[b!]
\includegraphics[width=1\columnwidth]{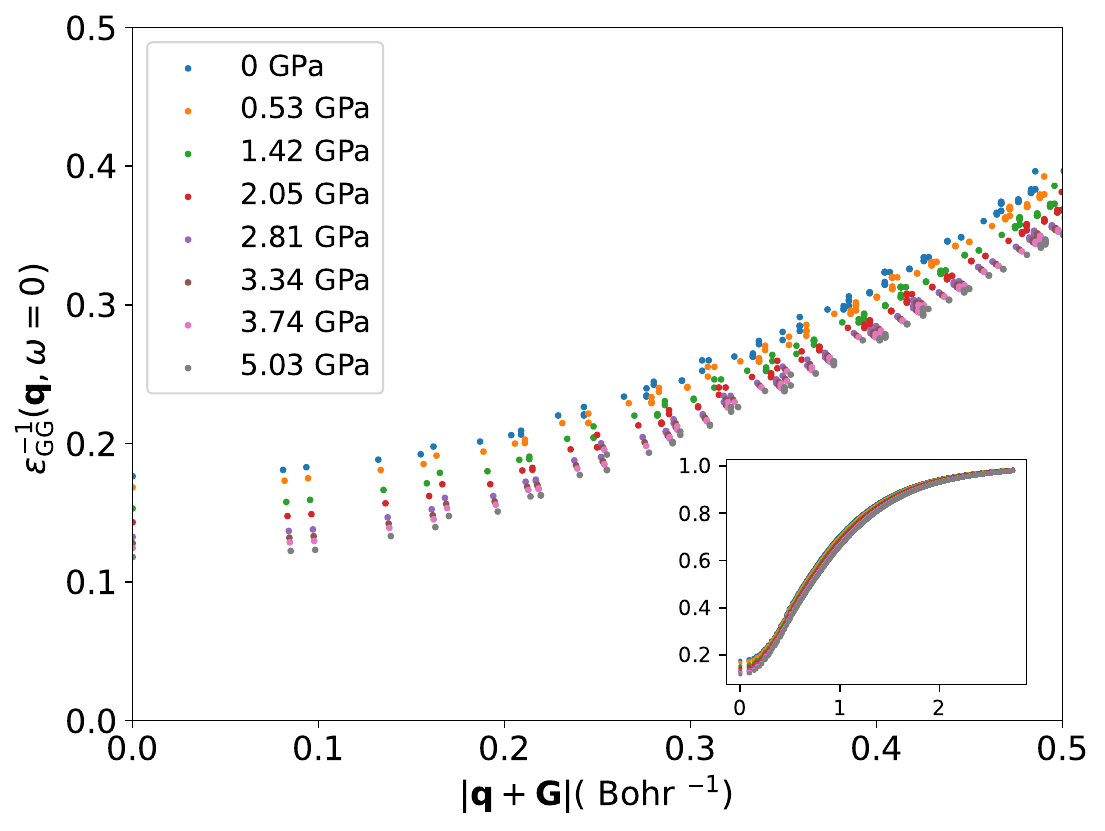}
    \caption{(Color online) Diagonal terms of of the inverse dielectric matrix for the all $\bf q$ points $\epsilon_{\mathbf{G G}}^{-1}\left(\mathbf{q}, \omega\right)$  at the first frequency $\omega$ = 0. The horizontal axis is limited within the range of 0 to 0.5 Bohr$^{-1}$, where it represents $|\mathbf{q}+\mathbf{G}|$. (Inset): The horizontal axis represents the entire range of $|\mathbf{q}+\mathbf{G}|$. }
\label{dielectric matrix}
\end{figure}

\begin{figure}[b!]
\includegraphics[width=1\columnwidth]{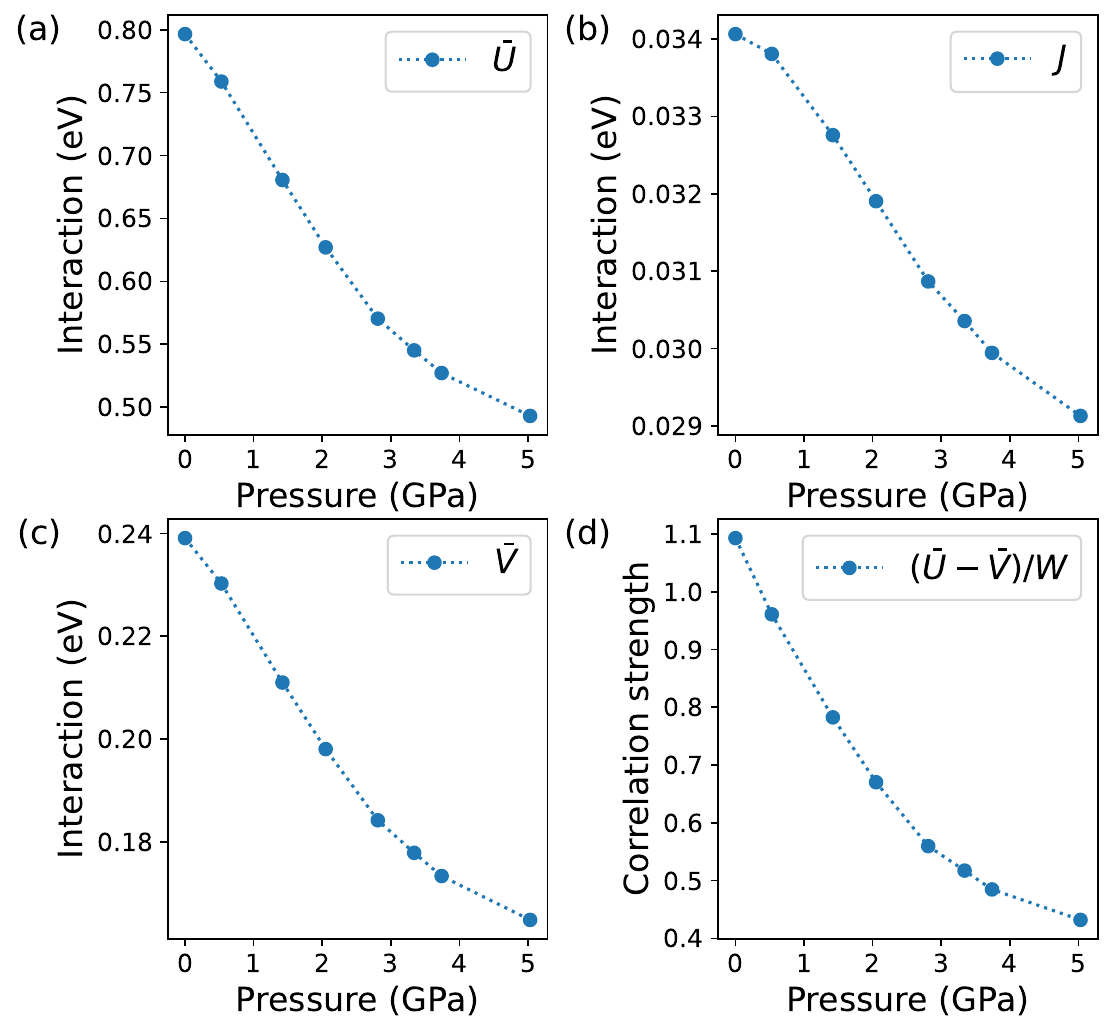}
    \caption{(Color online) Pressure dependence of (a) the average of
the on-site effective Coulomb repulsion $\bar{U}$, (b) the on-site effective
exchange interaction $J$, (c) the average of the off-site effective
Coulomb repulsion between neighboring sites $\bar{V}$, and (d) the correlation strength $(\bar{U} - \bar{V}) / W$, which are derived within the cRPA method.}
\label{Coulomb interaction}
\end{figure}

In this section, we analyze the effective Coulomb interaction parameters for K$_3$C$_{60}$. Specifically, we focus on the parameters $U = U_{ii}(\mathbf{0})$, $U^{\prime} = U_{ij}(\mathbf{0})$, and $J = J_{ij}(\mathbf{0})$, where $i \neq j$, $U_{i j}(\mathbf{R})$ in Eq.~(\ref{U}) and $J_{i j}(\mathbf{R})$ in Eq.~(\ref{J}).
$U$ is the intraorbital Coulomb interaction, $U^{\prime}$ is the interorbital interaction and $J$ is the exchange interactions.
Fig.~\ref{cRPA} shows our calculated interaction parameters $U$, $U^{\prime}$ and $J$ with two screening levels (bare and cRPA).
It can be observed that the values of effective interaction parameters decrease due to the screening process. For instance, at 0 GPa, the bare Coulomb repulsion for K$_3$C$_{60}$ is approximately 3.32 eV, but considering the cRPA screening effects, this value decreases to around 0.83 eV. 

Fig.~\ref{dielectric matrix} shows the inverse dielectric matrix $\left\{\epsilon_{\mathrm{{\bf GG}}}^{-1}(\mathbf{q}, \omega)\right\}$ and the macroscopic dielectric constant was defined as
\begin{align}
\epsilon _{M}^{\mathrm{cRPA}} = \lim_{\mathbf{Q} \rightarrow 0, \omega \rightarrow 0} \frac{1}{\epsilon _{\mathbf{GG}}^{\mathrm{cRPA}^{-1}}(\mathbf{q}, \omega)},
\label{epsilon}
\end{align}
where $\omega$ is the frequency, $\mathbf{Q} = \mathbf{q} + \mathbf{G}$, with $\mathbf{q}$ being the wave vector in the first Brillouin zone, and $\mathbf{G}$ being the reciprocal lattice vector. 
Fig.~\ref{cRPA}(d) present our calculated cRPA-macroscopic-dielectric constant $\epsilon_M^{\mathrm{cRPA}}$ in Eq.~(\ref{epsilon}).
The diagonal elements of the inverse dielectric matrix demonstrate lower values with increasing pressure (Fig.~\ref{dielectric matrix}), eventually converging to 1 with increasing $|\mathbf{q}+\mathbf{G}|$ (Fig.~\ref{dielectric matrix}(Inset)).

Fig.~\ref{Coulomb interaction} summarizes the results of the cRPA calculations: the on-site Coulomb repulsion $\bar{U}$ averaged over the MLWOs derived from the target band, the on-site exchange interaction $J$, the off-site interaction $\bar{V}$ averaged over the nearest-neighbor sites, and the ratio $(\bar{U}-\bar{V}) / W$ which measures the correlation strength in the system. Note that the net interaction is estimated as $\bar{U}-\bar{V}$, based on the analysis of the extended Hubbard model. 
Throughout the entire process, the relationship $U^{\prime} \sim U - 2J$ remains valid. As pressure increases, all parameters exhibit varying degrees of reduction, with the on-site Coulomb repulsion being notably affected by pressure. Specifically, $\bar{U}$ decreases from 0.80 eV to 0.50 eV, representing a decrease of approximately 40\%. The value of $J$ undergoes a relatively minor change, decreasing from 0.034 eV to 0.029 eV, corresponding to a decrease of approximately 15\%. Additionally, $\bar{V}$ decreases from 0.24 eV to 0.16 eV, approximately declining by 31\%. 
$\bar{U}$ is approximately three times $\bar{V}$ and twenty times $J$. $\bar{U}$ and $\bar{V}$ decrease almost synchronously with increasing pressure, while the rate of decrease in $J$ with pressure is slower than that of $\bar{U}$ and $\bar{V}$. In terms of ratios, $\bar{U}$/$J$ decreases from 23.4 to 17.0, while $\bar{V}$/$J$ only decreases from 7.0 to 5.7. 

In general, a small value of $J$ tends to favor low-spin states, consistent with experimental observations. It is worth mentioning that there is a proposal suggesting that the Jahn-Teller coupling may exert greater influence than the Hund's rule coupling $J$, and when coupled with a sufficiently large $U$, lead to the emergence of superconductivity. We will discuss this pertinent topic in Sec.~\ref{discussion}.
The ratio $(\bar{U} - \bar{V}) / W$, which measures the correlation strength in the system, shows simple monotonic decrease with applying pressure. 
It is noteworthy that at 0 GPa, the value of $(\bar{U} - \bar{V}) / W$ is approximately 1, suggesting the potential for C$_{60}$ superconductors to be strongly correlated electronic systems.
The results at 0 GPa are consistent with the $(\bar{U} - \bar{V}) / W \sim 1$~\cite{PhysRevB.85.155452} exhibited by other C$_{60}$ superconductors.
However, as pressure increases, this value gradually decreases, indicating a departure from a strongly correlated electronic system.

\subsection{Electron-Phonon Interactions and Phonon Frequencies}

The phonon modes associated with K$_3$C$_{60}$ that have the potential to interact with the $t_{1u}$ electrons can be categorized into distinct classes, including libration modes, intermolecular modes, optical modes involving alkali cations and C$_{60}^{3-}$ anions, and intramolecular modes. It has been demonstrated that, among these modes, the interactions of libration, intermolecular, and alkali-ion modes with the $t_{1u}$ electrons are relatively weak in comparison to those of the intramolecular modes~\cite{othermode1,othermode2,othermode3,othermode4,othermode5,othermode6}. Therefore, our primary focus lies on investigating the characteristics of intramolecular phonons.

Given the high symmetry of the C$_{60}$ molecule, only specific types of intramolecular phonon modes can couple with the $t_{1u}$ electrons due to symmetry considerations. If we consider the ideal icosahedral symmetry of the isolated C$_{60}$ molecule, a total of $60 \times 3 - 6 = 174$ intramolecular vibrational modes exist (where the subtraction of six modes pertains to the molecule's translational and rotational degrees of freedom). These modes are further categorized into two $A_{g}$ modes, one $A_{u}$ mode, three $T_{1g}$ modes, four $T_{1u}$ modes, four $T_{3g}$ modes, five $T_{3u}$ modes, six $G_{g}$ modes, six $G_{u}$ modes, eight $H_{g}$ modes, and seven $H_{u}$ modes.

The $A_{g(u)}$ modes exhibit no degeneracy, whereas the $T_{1 g(u)}$ and $T_{3 g(u)}$ modes, the $G_{g(u)}$ modes, and the $H_{g(u)}$ modes possess three-fold, four-fold, and five-fold degeneracies, respectively. Among the intramolecular modes, only the phonon modes with $A_g$ and $H_g$ symmetries exhibit finite electron-phonon couplings to the $t_{1u}$ electrons~\cite{doi:10.1126/science.254.5034.989,PhysRevB.44.12106}.

\begin{figure}[htb!]
\begin{center}
    \centering
    \includegraphics[width=\columnwidth]{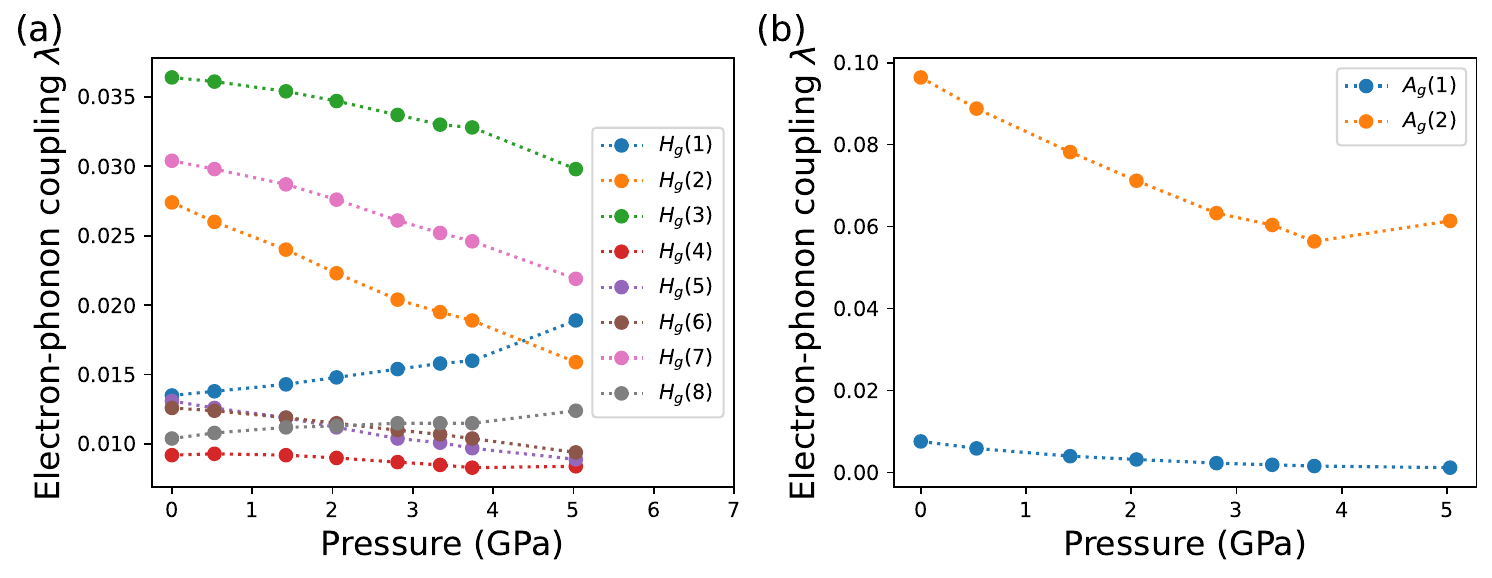}
    \caption{(Color online) Partially renormalized electron-phonon coupling parameter of different phonon modes at $\Gamma$ point calculated by cDFPT method. The phonon modes are as follows: (a) $H_g$(1)$\sim$(8), (b) $A_g$(1) and $A_g$(2).}
\label{phonon coupling}
\end{center}
\end{figure}

\begin{figure}[htb!]
\includegraphics[width=0.8\columnwidth]{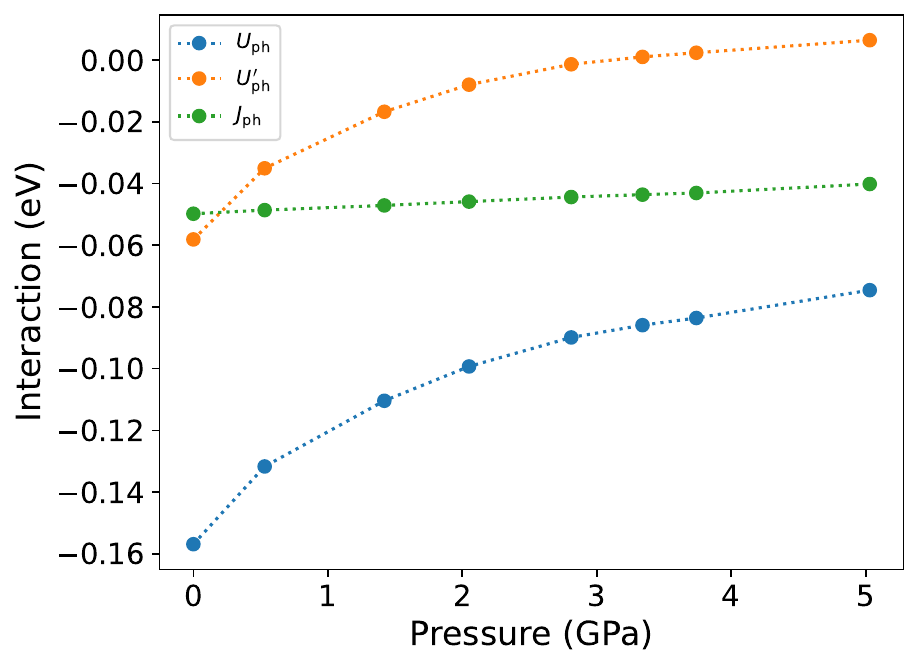}
    \caption{(Color online) Pressure dependence of effective phonon-mediated interaction.
    $U_{\rm ph}$ is the intraorbital density-density-type interaction, $U^{\prime}_{\rm ph}$ is the interorbital density-density-type interaction, $J_{\rm ph}$ is the exchange-type interaction, respectively.}
\label{phonon interaction}
\end{figure}

\begin{table*}[!ht]
\renewcommand{\arraystretch}{1.5}
    \caption{Partially renormalized phonon frequencies of different modes at $\Gamma$ point calculated by cDFPT method. The unit is in $\mathrm{cm}^{-1}\left(1\ \mathrm{eV}=8065.54 \mathrm{~cm}^{-1}\right)$.}
    \centering
    \resizebox{0.75\textwidth}{!}{%
    \begin{tabular}{clllllllllllllllllllllll}
    \hline\hline
     \multicolumn{1}{c}{Pressure (GPa)} & \multicolumn{1}{c}{$H_g$(1)} & \multicolumn{1}{c}{$H_g$(2)} & \multicolumn{1}{c}{$H_g$(3)} & \multicolumn{1}{c}{$H_g$(4)} & \multicolumn{1}{c}{$H_g$(5)} & \multicolumn{1}{c}{$H_g$(6)} & \multicolumn{1}{c}{$H_g$(7)} & \multicolumn{1}{c}{$H_g$(8)} & \multicolumn{1}{c}{$A_g$(1)} & \multicolumn{1}{c}{$A_g$(2)}  \\
        \hline
     0 & 259.41  & 427.28  & 685.73  & 779.26  & 1114.59  & 1273.56  & 1402.96  & 1531.60  & 504.82  & 1503.33    \\
    0.53 & 260.77  & 429.27  & 685.52  & 781.82  & 1117.53  & 1277.27  & 1407.12  & 1534.45  & 509.91  & 1504.94   \\
    1.42 & 262.92  & 432.06  & 685.25  & 785.49  & 1121.59  & 1282.41  & 1413.03  & 1538.64  & 514.54  & 1507.33    \\
    2.05 & 264.71  & 434.33  & 684.98  & 788.66  & 1124.99  & 1287.02  & 1418.17  & 1542.50  & 518.17  & 1509.58   \\
    2.81 & 266.85  & 437.14  & 684.52  & 792.61  & 1129.16  & 1292.71  & 1424.41  & 1547.37  & 522.46  & 1512.30    \\
    3.34 & 267.89  & 438.62  & 684.22  & 794.29  & 1130.64  & 1294.10  & 1427.68  & 1549.32  & 525.02  & 1514.18    \\
    3.74 & 268.91  & 439.71  & 684.05  & 796.28  & 1132.94  & 1297.92  & 1430.22  & 1552.10  & 527.33  & 1515.02    \\
    5.03 & 271.29  & 450.40  & 686.52  & 777.20  & 1104.26  & 1256.63  & 1374.34  & 1506.26  & 520.53  & 1453.92    \\
    \hline\hline   
        \end{tabular}%
    }
\label{phonon frequencies}
\end{table*}

Table~\ref{phonon frequencies} summarizes our calculated partially renormalized phonon frequencies of the all modes at $\Gamma$ point under different pressures.
The high phonon frequencies up to $\sim 1500 \mathrm{~cm}^{-1}(\sim 0.2$ $\rm {eV})$ can be ascribed to the stiff C$-\mathrm{C}$ bonds and the lightness of the carbon atoms. 

Note that these frequencies are inputs for the low-energy solvers and thus can not be directly compared with the experimentally observed frequencies. In general, the electron-phonon coupling of the individual mode is not large (see Fig.~\ref{phonon coupling}), while the accumulation of the contributions leads to the total electron-phonon coupling of $\lambda \sim 0.5-1.0$~\cite{Gunnarsson2004-xi}. 

In Fig.~\ref{phonon coupling}, we can visually observe the trends of various phonon modes as they change with pressure. With increasing pressure, the majority of phonon modes exhibit a decreasing trend in their intensities, with the exception of the $H_g$(1) and $H_g$(8) modes. This observed trend aligns with the overall decrease in phonon coupling strengths as the pressure increases.

Fig.~\ref{phonon interaction} summarizes the values of the static parts. We also find that the relation $U_{\rm ph}^{\prime} \sim U_{\rm ph}-2 J_{\rm ph}$ holds well. The negative values of $U_{\rm ph}$, $U_{\rm ph}^{\prime}$, and $J_{\rm ph}$ indicate that the interactions are attractive. 
Therefore, they will compete with the repulsive onsite Coulomb interactions. 
It is evident that an increase in pressure leads to a reduction in the absolute magnitudes of $U_{\rm ph}$ and $J_{\rm ph}$. The weakening trend is consistent for intra-orbital density-density-type interactions and exchange-type interactions. Additionally, the magnitude of $U_{\rm ph}^{\prime}$ decreases to approximately 3 GPa before undergoing a sign reversal, transitioning from negative to positive. This sign change indicates a gradual transition from attraction to repulsion in the inter-orbital density-density-type interactions.
As for the density-density channel, since the Coulomb interaction $U$ for the $t_{1 u}$ electrons (see Fig.~\ref{Coulomb interaction}(a)) is approximately five times larger in magnitude than electron-phonon coupling $U_{\rm ph} $, the repulsive Coulomb interaction dominates over the phonon mediated attraction.
However, remarkably, the situation changes for the exchange type interaction: at 0 GPa the absolute values of $\left|J_{\rm ph}\right| \sim 0.05$ $\mathrm{eV}$ are larger than those of the Hund's coupling $J \sim 0.034$ $\mathrm{eV}$ (see Fig.~\ref{Coulomb interaction}(b)). It indicates that, effectively, negative exchange interactions are realized. Thus the systems will obey the inverted Hund's rule, which favors the low-spin state rather than the high-spin state. This fact naturally explains the origin of the low-spin state observed in the Mott-insulating phase~\cite{Ganin2010-qd}. 

On the other hand, the enhancement of the negative $J_{\rm ph}$ due to the strong coupling between the Jahn-Teller modes and the $t_{1 u}$ electrons. As is already mentioned in the discussion above, the Jahn-Teller $H_g$ modes are the non-density-type electron-phonon coupling and  $A_g$ modes are density-type electron-phonon coupling.
It shows the $H_g$ modes contributes to $J_{\rm ph}$, the $A_g$ modes do not contribute.
We focus our attention on the behavior of the $H_g$ and $A_g$ modes, as they play key roles in these interactions. Notably, $A_g$ modes predominantly contribute to $| U_{\rm ph} |$, and their trends exhibit substantial similarity. Conversely, $H_g$ modes contribute primarily to $| J_{\rm ph} |$, and it is evident that the trends of $H_g$(2)$\sim$$H_g$(7) modes are largely consistent with $| J_{\rm ph} |$, although the intensities of $H_g$(1) and $H_g$(8) modes increase as pressure decreases.
Given that increasing pressure does not alter the relationship between $U_{\rm ph}^{\prime} $ and $ U_{\rm ph}-2 J_{\rm ph}$, it can be inferred that the contributions of $H_g$(1) and $H_g$(8) modes become more prominent as pressure increases compared to other modes. This leads to a relatively smaller reduction in $J_{\rm ph}$ compared to $U_{\rm ph}$, resulting in the reversal of the inter-orbital density-density-type interaction $|U_{\rm ph}^{\prime}|$ from attraction to repulsion.

\subsection{Effective interactions}
\label{effective interactions}

\begin{figure}[b!] 
\includegraphics[width=1\columnwidth]{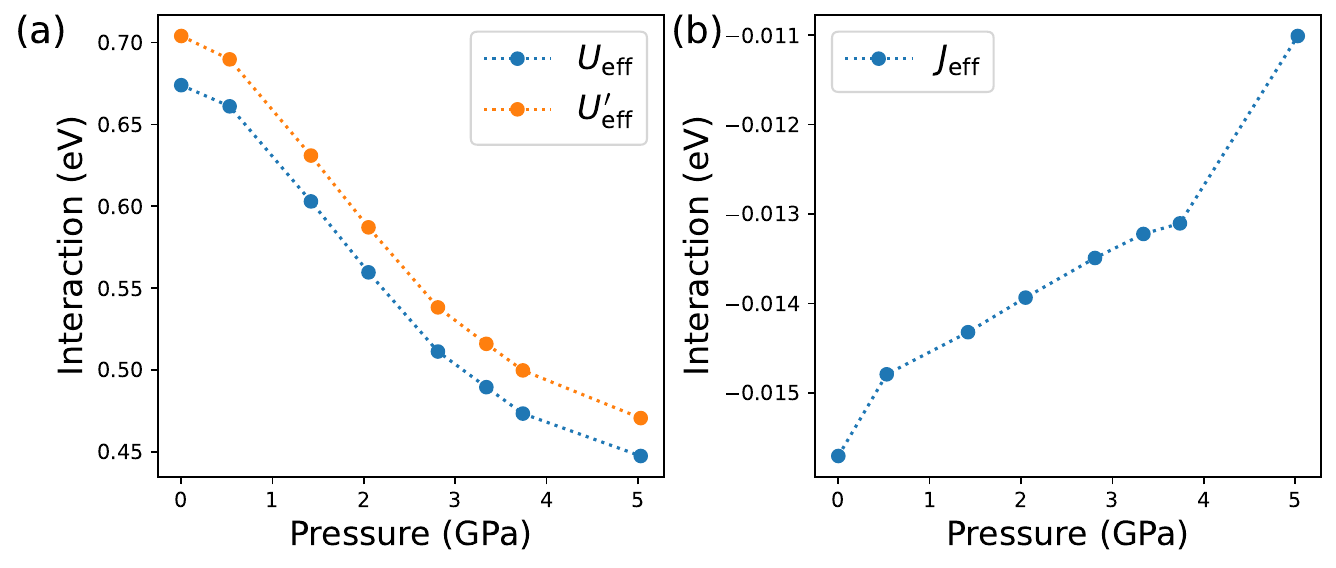}
    \caption{(Color online) Pressure dependence of effective interaction $U_{\rm eff}(U+U_{\rm ph})$, $U^{\prime}_{\rm eff}(U^{\prime}+U^{\prime}_{\rm ph})$ and $J_{\rm eff}(J+J_{\rm ph})$. $ U[U_{\rm ph}]$, $U^{\prime}\left[U^{\prime}_{\rm ph}\right]$, and $J\left[J_{\mathrm{ph}}\right]$ are the intraorbital, interorbital, and exchange components of the cRPA Coulomb [cDFPT phonon-mediated] interactions, respectively. }
\label{Pressure independence of effective interaction}
\end{figure}

\begin{figure}[b!]

\includegraphics[width=1\columnwidth]{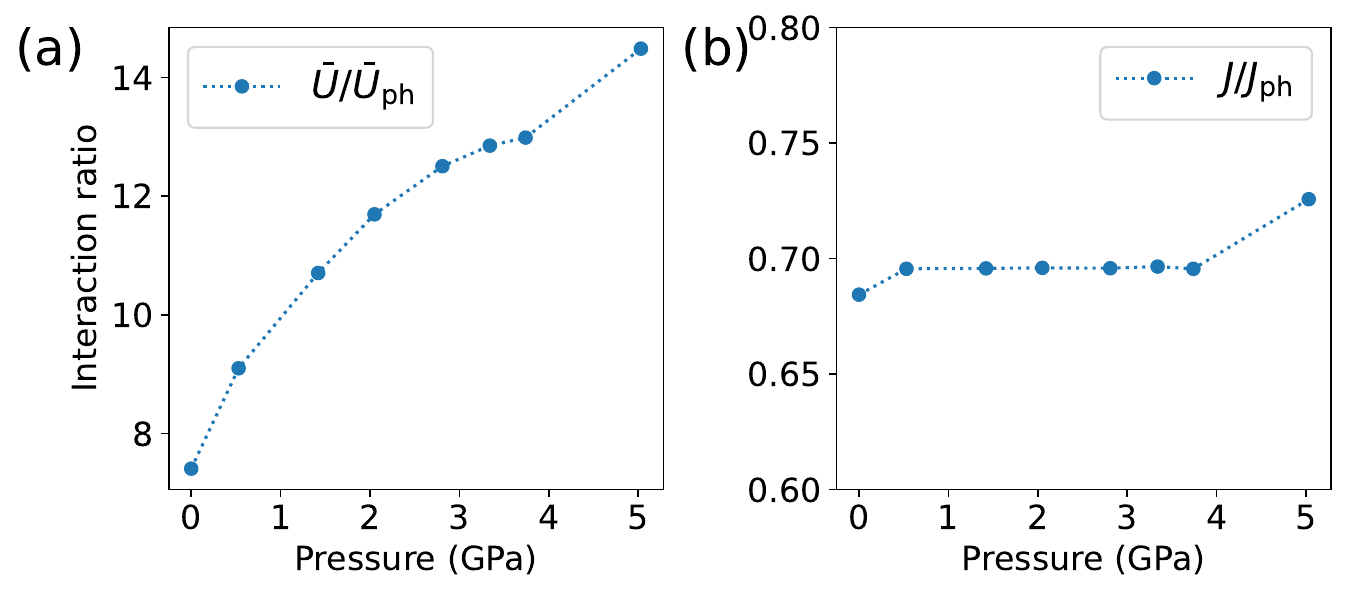}
    \caption{(Color online) Pressure dependence of the comparison of effective interaction. $ \bar{U}[\bar{U}_{\rm ph}]$,  and $J\left[J_{\mathrm{ph}}\right]$ are the on-site effective Coulomb repulsion and exchange components of the cRPA Coulomb [cDFPT phonon-mediated] interactions, respectively.}
\label{compareing effective interaction}
\end{figure}
Fig.~\ref{Pressure independence of effective interaction} shows the pressure dependence of the effective interactions between the $t_{1u}$ electrons ($U_{\rm eff}=U+U_{\rm ph}$,  $U^{\prime}_{\rm eff}=U^{\prime}+U^{\prime}_{\rm ph}$,  $J_{\rm eff}=J+J_{\rm ph}$), which are given by the sum of the Coulomb $U$, $U^{\prime}$, $J$ and the retarded phonon-mediated interactions $U_{\rm ph}$, $U^{\prime}_{\rm ph}$ and $J_{\rm ph}$.
In the context of K$_3$C$_{60}$ subjected to variable external pressure conditions, several notable observations have emerged. Firstly, it has been observed that both $U_{\rm eff}$ and $U^{\prime}_{\rm eff}$ remain positive across the entire range of pressures studied, exhibiting substantial variations as their values decrease from 0.7 eV to 0.45 eV. Notably, these two parameters exhibit a highly synchronized trend, demonstrating a strong degree of synchronicity.
Secondly, the effective exchange interaction parameter, $J_{\rm eff}$, consistently takes negative values under all pressure conditions, with its variation confined to a narrow range, specifically within the limits of $-$0.016 eV to $-$0.011 eV. It is pertinent to note that $U_{\rm eff}$ and $U^{\prime}_{\rm eff}$ exhibit pronounced repulsive tendencies throughout the studied pressure range, which are greater than the attractive tendencies exhibited by $J_{\rm eff}$.

Furthermore, the transition of $J_{\rm eff}$ to negative values is attributed to the contributions from phonon. Within the parameter regime where $J_{\rm eff}$ becomes negative, as $U^{\prime}_{\rm eff}$ is observed to marginally surpass $U_{\rm eff}$, indicating that the interorbital repulsion is slightly greater than the intraorbital repulsion across all pressure conditions. This scenario favors electron pairing within the orbitals, consistent with the conclusions drawn regarding intraorbital pairs~\cite{Naren2008-kd,Capone2009-uq,Capone2001-ip,Capone2004-nn,Nomura2015-kz}.

Fig.~\ref{compareing effective interaction} illustrates the pressure-induced variations in the strengths of the Coulomb interaction and the electron-phonon interaction parameters. In particular, it compares the behavior of on-site effective interaction ($\bar{U}$) with that of the phonon-mediated interaction ($U_{\rm ph}$). Notably, $\bar{U}$ exhibits a slower rate of reduction compared to $U_{\rm ph}$. This implies that as pressure increases, the intraorbital electron repulsion decreases at a slower rate than the intraorbital electron-phonon attraction.
On the other hand, the exchange interaction parameter ($J$) exhibits a lower sensitivity to pressure variations. Unlike $\bar{U}$, it does not display as pronounced changes with increasing pressure.

\section{discussion}
\label{discussion}

We have carried out a fully $ab$ $initio$ study of all the interactions of a multiorbital Hubbard model coupled with phonons are computed from first principles with only the information on the atomic positions. 
In our study, we unveil changes in electronic and phononic properties induced by pressure fluctuations and compare several explanations for the unconventional superconducting mechanism in K$_3$C$_{60}$. Referring to the experimental trend of decreasing  $T_c$ in K$_3$C$_{60}$ as pressure increases~\cite{Wang2022-jd,Sparn1991-kz}, we make the following observations:

1) As pressure increases, the strength of the Coulomb interaction consistently surpasses that of the electron-phonon interaction, albeit both exhibit decreasing trends. 
Furthermore, phonon mediated intaraction decrease in strength more rapidly than Coulomb interaction. The concurrent weakening of the superconducting phase and phonon interactions aligns well with the proposition that local pairing with Jahn-Teller phonons assisted by Coulomb repulsion predominates as the superconducting mechanism~\cite{Nomura2016-lk,Gunnarsson1995-qr,Han2003-fv} in K$_3$C$_{60}$. This highlights the impact of phonon interactions on the superconducting mechanism, even though the phonon interaction strength in K$_3$C$_{60}$ is relatively modest, due to the dominant role of strong Coulomb interactions in suppressing kinetic energy.

2) Through band structure calculations, we assert that the primary source of electrons within K$_3$C$_{60}$ relies on the $t_{1u}$ band. However, the density of states plot in Fig.~\ref{dos} reveals that the $t_{1u}$ band is entirely contributed by electrons from the carbon atoms, independent of potassium atoms. This observation indicates that the hypothesis suggesting that potassium ion phonon modes exert a strong attractive force on electrons on the C$_{60}$, sufficient to overcome Coulomb repulsion and induce superconductivity~\cite{Zhang1991-jc}, is incorrect. The role of doping potassium ions merely transitions C$_{60}$ from an insulator to a metal but has minimal influence on the superconducting mechanism.

3) In K$_3$C$_{60}$, when considering only the Coulomb interaction in isolation, we observe that $U$ is greater than $U^{\prime}$. However, upon incorporating electron-phonon interactions, denoted as $U_{\rm ph}$ and $U^{\prime}_{\rm ph}$, we discern that $U_{\rm eff}$ becomes smaller than $U^{\prime}_{\rm eff}$. This observation forms the foundational explanation for electron pairing. Despite the relatively modest strength of phonon interactions within K$_3$C$_{60}$, the robust Coulomb interactions effectively curtail the kinetic energy of electrons. 
This, in turn, facilitates the interorbital tunneling of electron pairs through pair-hopping interactions, which is an indispensable mechanism for the emergence of superconductivity.
Throughout the entire process, both $U_{\rm eff}$ and $U^{\prime}_{\rm eff}$ decrease with increasing pressure, with $U_{\rm eff}$ consistently being less than $U^{\prime}_{\rm eff}$. This indicates that interorbital Coulomb interactions are greater than intraorbital interactions, implying that electron pairing within the orbitals persists even as Coulomb interactions decrease.
Furthermore, the $A_g$ mode give $U_{\rm ph}$ = $U^{\prime}_{\rm ph}$, and $H_g$ mode give negative $J_{\rm ph}$ and negative $J_{\rm eff}$.
Hence $H_g$ modes assume a pivotal role in determining the relationship where $U^{\prime}_{\rm eff}$ = $U_{\rm eff} - 2 J_{\rm eff} > U_{\rm eff}$,
both modes are indispensable in shaping the dynamics of the superconducting mechanism.

\section{Conlusion}
\label{Conlusion}

In conclusion, our comprehensive $ab$ $initio$ study of a multiorbital Hubbard model coupled with phonons under varying pressure conditions sheds light on the intriguing behavior of the superconducting mechanism in K$_3$C$_{60}$. Under increasing pressure, we observe significant changes in the system's parameters and theoretical concepts. As pressure increases, we observe that despite the relatively smaller strength of phonon interactions, their impact on the superconducting mechanism in K$_3$C$_{60}$ is significant. This is due to the local pairing with Jahn-Teller phonons~\cite{Nomura2016-lk,Gunnarsson1995-qr,Han2003-fv}, which is further assisted by the suppressive effect of strong Coulomb interactions on kinetic energy.

Furthermore, our study reveals that interorbital Coulomb interactions remain greater than intraorbital interactions throughout the pressure increase, indicating that electron pairing within the orbitals persists despite decreasing Coulomb interactions. 
Pressure-induced changes in the electronic and phononic properties of the system, along with variations in key parameters like $U_{\rm eff}$, $U^{\prime}_{\rm eff}$ and $J_{\rm eff}$, highlight the intricate interplay between different factors in the superconducting mechanism of K$_3$C$_{60}$. This study advances our understanding of unconventional superconductivity in complex materials and underscores the importance of considering both electronic and phononic contributions in such systems.

\section*{Acknowledgement}
This work is supported by the National Natural Science Foundation of China~(Grant No.  12204130), Shenzhen Start-Up Research Funds~(Grant No. HA11409065), the HITSZ Start-Up Funds~(Grant No. X2022000). Yusuke Nomura acknowledges the support by JSPS KAKENHI (No. JP23H04869 and JP21H01041). Haipeng Li acknowledges the support by the Fundamental Research Funds for the Central Universities of China (Grant No. 2019ZDPY16) and Basic Research Project of Xuzhou City (Grant No. KC22043).





\bibliographystyle{apsrev4-1.bst}
\bibliography{references}
\end{document}